\documentclass[twocolumn,prl,showpacs,superscriptaddress,showpacs]{revtex4-1}
\usepackage{graphicx}
\usepackage{amssymb,amsmath}
\usepackage{bm}
\usepackage{dcolumn}
\usepackage{subfigure}
\usepackage{float}
\usepackage{url}
\usepackage{color}
\usepackage{txfonts}
\usepackage[dvipdfmx]{hyperref}
\hypersetup{backref,pdfpagemode=FullScreen,colorlinks=true,breaklinks,urlcolor=blue,linkcolor=blue,citecolor=blue}

\begin{document}
\title{Dynamical Generation of Topological Magnetic Lattices for Ultracold Atoms}

\affiliation{State Key Laboratory of Low Dimensional Quantum Physics, Department of Physics, Tsinghua University, Beijing 100084, China}
\affiliation{MOE Key Laboratory of Fundamental Physical Quantities Measurements, School of Physics, Huazhong University of Science and Technology, Wuhan 430074, China}
\affiliation{Department of Physics and Astronomy, University of Pittsburgh, Pittsburgh, Pennsylvania 15260, USA}
\affiliation{Collaborative Innovation Center of Quantum Matter, Beijing 100084, China}

\author{Jinlong Yu}
\affiliation{State Key Laboratory of Low Dimensional Quantum Physics, Department of Physics, Tsinghua University, Beijing 100084, China}
\affiliation{Department of Physics and Astronomy, University of Pittsburgh, Pittsburgh, Pennsylvania 15260, USA}
\author{Zhi-Fang Xu}
\email{zfxu83@gmail.com}
\affiliation{MOE Key Laboratory of Fundamental Physical Quantities Measurements, School of Physics, Huazhong University of Science and Technology, Wuhan 430074, China}
\affiliation{Department of Physics and Astronomy, University of Pittsburgh, Pittsburgh, Pennsylvania 15260, USA}
\author{Rong L\"u}
\affiliation{State Key Laboratory of Low Dimensional Quantum Physics, Department of Physics, Tsinghua University, Beijing 100084, China}
\affiliation{Collaborative Innovation Center of Quantum Matter, Beijing 100084, China}
\author{Li You}
\affiliation{State Key Laboratory of Low Dimensional Quantum Physics, Department of Physics, Tsinghua University, Beijing 100084, China}
\affiliation{Collaborative Innovation Center of Quantum Matter, Beijing 100084, China}

\begin{abstract}
  We propose a scheme to dynamically synthesize a space-periodic effective magnetic field
  for neutral atoms by time-periodic magnetic field pulses.
    When atomic spin adiabatically follows the direction of the effective magnetic field,
an adiabatic scalar potential together with a geometric vector potential emerges for the atomic center-of-mass
motion, due to the Berry phase effect.
While atoms hop between honeycomb lattice sites formed by the minima of the adiabatic potential,
complex Peierls phase factors in the hopping coefficients are induced by the vector potential, which facilitate a topological Chern insulator.
  With further tuning of external parameters, both a topological phase transition and topological flat bands can be achieved, highlighting realistic prospects for studying strongly correlated phenomena in this system.
Our Letter presents an alternative pathway towards creating and manipulating topological states of ultracold atoms by magnetic fields.
\end{abstract}


\maketitle

Gauge fields lie at the center of our modern
understanding of physics in many systems, including those of high energy and condensed matter, as well as of ultracold atoms. Within the gauge-field 
paradigm, interactions between particles, which enable rich quantum phases of a many-body system, are mediated through
gauge fields.
For instance,
solid-state materials with charged quasiparticles in magnetic fields or with spin-orbit coupling (SOC)
show a rich variety of quantum Hall effects and exotic topological superconductivity~\cite{Hasan2010,Qi2011}.
The interplay between gauge fields and lattice systems is also of great interest (for a pedagogical review, see~\cite{Bernevig2013}).
The spectrum of a charged particle in a square lattice exposed to a strong uniform magnetic field
shows {{a}} fractal structure, widely known as the Hofstadter butterfly~\cite{Hofstadter1976}.
In another seminal work, Haldane shows that the quantum Hall effect without Landau levels can be realized when a periodically staggered magnetic field is applied to charged particles in a honeycomb lattice~\cite{Haldane1988}.

Ultracold atoms in lattice systems are considered to be powerful simulators for studying gauge-field physics~\cite{Dalibard2011, Goldman2014b, Goldman2014, Struck_Sengstock2011, Hauke_Lewenstein2012, Jaksch2003, Celi-Lewenstein2014, Neupert2014, Zheng2014, Aidelsburger2013, Miyake2013, Aidelsburger2015, Ketterle2015, Mancini_Fallani2015, Stuhl_Spielman2015, Jotzu2014, Cooper2011, Jimenez2012}.
 Both the Hofstadter and the Haldane models with cold atoms
were theoretically proposed~\cite{Jaksch2003, Celi-Lewenstein2014, Neupert2014, Zheng2014} and
experimentally demonstrated~\cite{Aidelsburger2013, Miyake2013, Aidelsburger2015, Ketterle2015, Mancini_Fallani2015, Stuhl_Spielman2015, Jotzu2014} by
making use of novel forms of light-atom interactions~\cite{Goldman2014b},
such as laser-assisted tunneling~\cite{Aidelsburger2013, Miyake2013, Aidelsburger2015, Ketterle2015}, the shaking-optical-lattice technique~\cite{Jotzu2014}, and SOC within a synthetic dimension~\cite{Mancini_Fallani2015, Stuhl_Spielman2015}.
In addition to the optical lattice formed from a space-periodic
ac-Stark shift
by interfering laser beams, proposals for the generation of a magnetic lattice with a space-periodic Zeeman shift
have been put forth~\cite{Yin2002, Zimmermann2005, Singh2008, Whitlock2009, Jose2014, Grabowski2003, Zimmermann2005, Singh2008, Whitlock2009, Leung2011, Jose2014, Romero-Isart2013, Luo2014} (and some have been realized~\cite{Zimmermann2005, Singh2008, Whitlock2009, Jose2014}) using current-carrying wires~\cite{Yin2002}, microfabricated wires or
permanent magnetic structures on atomic chips~\cite{Grabowski2003, Zimmermann2005, Singh2008, Whitlock2009, Leung2011, Jose2014}, superconducting vortex lattice shields~\cite{Romero-Isart2013}, and phase imprinting by gradient magnetic pulses~\cite{Luo2014}. In contrast to optical lattices,
magnetic lattices are free from atomic spontaneous emissions that are always accompanied by heating and loss of atoms. Additionally, they have the potential to reach shorter lattice constants~\cite{Leung2011, Romero-Isart2013} (of a few tens of nanometers as proposed in Ref.~\cite{Romero-Isart2013}), leading to improved energy scales
and less stringent temperature requirements; the lattice constants can even be continuously tuned~\cite{Luo2014}.
These advantageous features enhance the performance of atomic quantum gases as powerful quantum simulators.

While the simulation of gauge-field physics and the manipulation of topological states in optical lattices have shown fruitful results~\cite{Dalibard2011, Goldman2014b, Goldman2014, Struck_Sengstock2011, Hauke_Lewenstein2012, Jaksch2003, Celi-Lewenstein2014, Neupert2014, Zheng2014, Aidelsburger2013, Miyake2013, Aidelsburger2015, Ketterle2015, Mancini_Fallani2015, Stuhl_Spielman2015, Jotzu2014, Cooper2011, Jimenez2012},
it remains to show whether this is also the case for magnetic lattices. This Letter
provides an affirmative first answer to this question.

This Letter presents a scheme for synthesizing a time-independent effective Hamiltonian
with nontrivial band topology for atomic gases with internal spin degrees of freedom,
 based on the phase imprinting technique~\cite{Burger1999, Denschlag2000}.
 A two-dimensional (2D) magnetic lattice with triangular geometry emerges in the effective Hamiltonian.
In the limit when an atom is confined in the lowest Zeeman level,
an adiabatic scalar potential and a geometric vector potential are simultaneously generated for the center-of-mass motion~\cite{Dalibard2011, Cooper2011, Jimenez2012}.
The adiabatic scalar potential surface can form a honeycomb lattice, while the associated geometric
vector potential provides complex {{phases}} for next-nearest-neighbor (NNN) hopping coefficients
in realizing the Haldane model~\cite{Haldane1988, Shao2008}.
With the flexibility and tunability of magnetic fields, our scheme can be extended to produce a set of effective Hamiltonians  whose lowest energy bands undertake a topological phase transition from a topological (Chern) insulator to a trivial one. Moreover, models possessing topological quasiflat bands are realized near the phase-transition point.

\emph{The protocol.}---We consider a pancake-shaped quasi-2D ultracold atomic gas of spin $F$ confined in the $x$-$y$ plane (at $z=0$).
In the presence of a bias magnetic field $ B_0 \hat e_z$, the single-particle Hamiltonian is given by
\begin{eqnarray}\label{free_H}
  H_0 = \frac{{\bf{p}}^2}{2m} + \hbar \omega_0 F_z,
\end{eqnarray}
where ${\bf{p}} = (p_x, p_y)$ is the 2D kinetic momentum, $m$ is the atomic mass,
$\hbar$ is the reduced Planck constant, $F_z$ is the third component of the atomic spin vector (in unit of $\hbar$) ${\mathbf F}=(F_x,F_y,F_z)$ and
$\omega_0 = g_F \mu_B B_0/\hbar$ is the Larmor frequency at $B_0$,
where $g_F$ is the Land$\acute{\text{e}}$ $g$ factor for the spin-$F$ hyperfine state manifold and $\mu_B$ is the Bohr magneton.

A short gradient-magnetic-field pulse ${{B'y\hat e_y}}$ of
duration $\delta t'$ imprints a space-dependent phase factor ~\cite{Burger1999, Denschlag2000, Xu2013, Anderson2013, Luo2014, Luo2015} onto the wave function
as $\exp \left( -{i{k_{\text {so}}}y{F_y}}\right)$,
where ${k_{{\text{so}}}}=\delta t'{g_F}{\mu _B}B'/\hbar$ is the SOC strength~\cite{Xu2013, Anderson2013} with $B'$ the magnetic gradient.
After a free evolution time $\delta t$, a second magnetic field pulse in the
opposite direction imprints an opposite phase.
The two pulses combined together enact a unitary transformation
\begin{eqnarray}\label{F_z_rotated_along_y}
  {e^{i{k_{\text{so}}}y{F_y}}}{F_z}{e^{ - i{k_{\text{so}}}y{F_y}}} = {F_z}\cos \left( {{k_{{\text{so}}}}y} \right) - {F_x}\sin \left( {{k_{{\text{so}}}}y} \right),
\end{eqnarray}
which rotates the magnetic field $B_0\left(0,0,1\right)$ to a
 space-periodic form $B_0(-\sin(k_\text{so}y), 0, \cos(k_\text{so}y) )$.
 Similarly, an opposite uniform-field pulse pair $\mp B_y \hat e_y$
 with a pulse area $\delta t'{g_F}{\mu _B}{B_{y}}/\hbar=\pi$
inverts the magnetic field $B_0\left(0,0,1\right)$ to $B_0\left(0,0,-1\right)$
as ${e^{-i\pi {F_y}}}{F_z}{e^{  i\pi {F_y}}} =  - {F_z}$.
More generally, a gradient magnetic field pulse along an arbitrary direction
$\hat e_\theta = \left(\cos\theta, \sin\theta, 0\right)$ in the $x$-$y$ plane imprints
a phase factor $\exp \left( -{i{k_{\text{so}}}r_\theta{F_\theta}} \right)$,
where ${r_\theta } = {\mathbf{r}} \cdot {{\hat e}_\theta } = x\cos \theta  + y\sin \theta $
and ${F_\theta } = {\mathbf{F}} \cdot {{\hat e}_\theta } = {F_x}\cos \theta  + {F_y}\sin \theta $
are, respectively, the coordinate vector $\mathbf{r} = \left( x, y \right)$ and
the spin vector ${\mathbf{F}}$ projected to the $\hat e_\theta$ direction.
Following a period of
free evolution and a second pulse from an opposite gradient field,
an expression analogous
to Eq.~(\ref{F_z_rotated_along_y}) generates a magnetic field with spatial periodicity
along the $\hat e_\theta$ direction.

\begin{figure}[tbp]
  \includegraphics[width=\columnwidth]{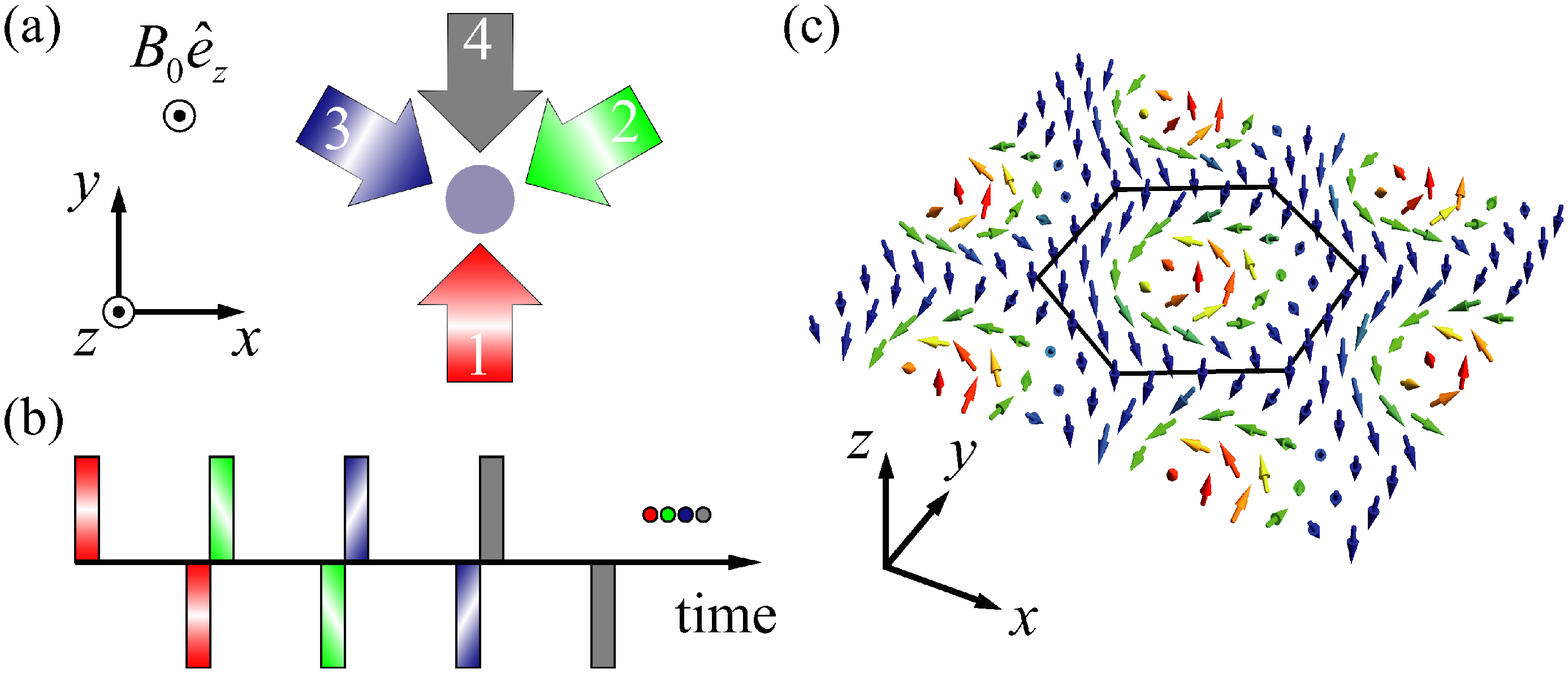}\\
  \caption{A schematic illustration for synthesizing a magnetic lattice.
  (a) An atomic cloud is exposed to a uniform bias magnetic field $B_0 \hat e_z$
  (pointing out of the page), and subjected in sequence to four pairs of opposite magnetic pulses as shown in (b).
  Pulse pairs $1$, $2$, and $3$ are gradient magnetic fields lying in the $x$-$y$ plane along directions separated by
  $120^\circ$. Pulse pair $4$ executes $\mp \pi$ spin rotations along the $y$ direction. The color gradient represents the corresponding magnetic field strength.
  (b) The time-periodic pulse sequence with four pairs of pulses forms one complete evolution period.
  (c) A three-dimensional view of the effective magnetic field $ {\mathbf{ B}}_{{\rm {eff}}}$ [Eq.~(\ref{B_eff})],
  which forms a Skyrmion lattice~\cite{Nagaosa2013} with its
   Wigner-Seitz unit cell shown by the hexagon [of edge length $4\pi/(3k_{ \text{so}})$].
  The arrows are colored by the magnitude of the third component of ${\mathbf{ B}}_{{\rm {eff}}}$.
   }\label{Fig_1}
\end{figure}

In our scheme, discussed below, repeated pulse pairs are concatenated.
A complete cycle of the evolution period contains three gradient pulse pairs along directions
separated by an angle of ${{120^\circ}}$, together with a ${{\mp}}\pi$ pulse pair along the $y$ direction  as shown in Fig.~\ref{Fig_1}(a)-\ref{Fig_1}(b).
The total evolution over one complete cycle (of period $T=4\delta t$) is then given by
\begin{eqnarray} \label{U_T}
  \begin{gathered}
  U\left( {T,0} \right) = {e^{- i\pi {F_y}}}{e^{ - i{H_0}\delta t/\hbar }}{e^{  i\pi {F_y}}} \hfill \\
  \quad  \times \prod\limits_{j = 3,2,1} {{e^{i{k_{\text{so}}}{r_{{\theta _j}}}{F_{{\theta _j}}}}}{e^{ - i{H_0}\delta t/\hbar }}{e^{ - i{k_{\text{so}}}{r_{{\theta _j}}}{F_{{\theta _j}}}}}},  \hfill \\
\end{gathered}
\end{eqnarray}
with $\theta_j = -{\pi}/{6} + {2\pi j}/{3}$. According to the Floquet theorem~\cite{Maricq1982, Slichter1990}, a time-independent effective Hamiltonian can be defined according to $U\left( {T,0} \right) \equiv \exp \left({ - i{H_{{\text{eff}}}}T/\hbar }\right)$.
To the lowest order of $T$, we find~\cite{SuppMat2015}
\begin{eqnarray}\label{H_eff}
  {H_{{\rm{eff}}}} = \frac{1}{{2m}}{\left( {{\mathbf{p}} -
  \frac{3}{8}\hbar {k_{{\text{so}}}}{{\mathbf{ F}}_\bot}} \right)^2} + \frac{{15}}{{64}}\hbar {\omega _{{\text{so}}}} {{{\bf {F}}_\bot^2}}
   +{g_F}{\mu _B}{{\mathbf{B}}_{{\text{eff}}}}\cdot{\mathbf{F}}, \hskip 12pt
\end{eqnarray}
where ${{\mathbf{ F}}_\bot} = (F_x, F_y)$ is the 2D spin operator,
${\omega _{ \text{so}}} = \hbar k_{ \text{so}}^2/2m $ is the SOC frequency,
and ${{\mathbf B}_{{\text{eff}}}}$ is an effective magnetic field whose three components are given by
\begin{eqnarray}\label{B_eff}
   \begin{aligned}
  {B_{{\rm eff},x}} &= - \frac{{{B_0}}}{4}\sum\nolimits_j {\sin \left( {{k_{\text{so}}}{r_{\theta_j}}} \right)\sin {\theta _j}},   \\
  {B_{{\rm eff},y}} &=  \frac{{{B_0}}}{4}\sum\nolimits_j {\sin \left( {{k_{\text{so}}}{r_{\theta_j}}} \right)\cos {\theta _j}},   \\
  {B_{{\rm eff},z}} &= \frac{{{B_0}}}{4}\left[ - 1 + {\sum\nolimits_j {\cos \left( {{k_{\text{so}}}{r_{\theta_j}}} \right)}  } \right]. \\
\end{aligned}
\end{eqnarray}
The first two terms in Eq.~(\ref{H_eff}) arise from the unitary transformations by gradient pulse pairs
applied to the momentum operator~\cite{Xu2013,Luo2014}. The third term describes a magnetic (Zeeman) lattice that couples the atomic spin to the effective magnetic field ${{\mathbf B}_{{\text{eff}}}}$, as shown in Fig.~\ref{Fig_1}(c).

\emph{Geometric potentials and energy spectrum.}---The above protocol
for the generation of a triangular magnetic lattice is general, and can be applied to
atoms with arbitrary spins.
For concreteness, we choose a specific atomic species, fermionic $^6$Li,
with electron spin $J=1/2$, nuclear spin $I = 1$, and we consider the total hyperfine spin $F= I-J = 1/2$ ground-state manifold.
The Land$\acute{\text{e}}$ $g$ factor can be evaluated according to the Breit-Rabi formula~\cite{Woodgate1980}
to be ${g_F} \approx  - 1/3$.
The spin operator reduces to $\mathbf{ F} = \bm{\sigma} /{2}$, where $\bm{\sigma}$ is the vector of Pauli matrices.
To be more specific, in all numerical calculations, we assume a set of fixed
parameters unless otherwise noted. They are
$B_0 = 20\,\rm{mG}$, $B'\delta t' = 2\,{\text{G}}\,{\text{cm}}^{ - 1}\,{\text{ms}}$~\cite{[{A magnetic gradient pulse with $B'$ as large as $400\,\text{kG}\,\text{cm}^{-1}$ and duration $\delta t'$ less than $ 1 \mu \text{s}$ is achivable in  state-of-the-art atomic chip experiments; see for example }] Machluf2013},
which correspond to $k_{\text{so}}=(1.7\,\mu \rm{m})^{-1}$
and $\omega_0 = 32.3\omega_{\text{so}}=(2\pi)\times9.3\,\rm{kHz}$
for the $F=1/2$ manifold of $^6$Li.
With these parameters,
the lattice term in Eq.~(\ref{H_eff}) dominates during time evolution.

We denote the space-dependent eigenstates of the magnetic lattice by $\left| {{\chi _ {1,2} }(\mathbf{ r})} \right\rangle $, which satisfy
\begin{eqnarray}
  {g_F}{\mu _B}{{\mathbf{B}}_{{\text{eff}}}} \cdot \frac{{\bm{\sigma }}}{2}\left| {{\chi _ {1,2} }({\mathbf{r}})} \right\rangle  =  \pm {\epsilon _0}({\mathbf{r}})\left| {{\chi _ {1,2} }({\mathbf{r}})} \right\rangle ,
\end{eqnarray}
where ${\epsilon _0}({\mathbf{r}}) =  - |{g_F}{\mu _B}{{\mathbf{B}}_{{\text{eff}}}}|/2$ is the adiabatic potential
for atomic center-of-mass motion in the lower-energy eigenstate $\left| {{\chi _ {1} } } \right\rangle$.
For an atom adiabatically moving in this space-periodic Zeeman level, a vector potential $\mathbf{A}(\mathbf{r})$  emerges~\cite{SuppMat2015, ScalarPotential2015},
\begin{eqnarray}
 {\mathbf{A}} = i\hbar \left\langle {\left. {{\chi _1}} \right|\nabla {\chi _1}} \right\rangle  + \frac{3}{{16}}\hbar {k_{{\text{so}}}}\left\langle {{\chi _1}} \right|{{\bm{ \sigma }}_\bot }\left| {{\chi _1}} \right\rangle,
\end{eqnarray}
with ${ {\bm{ \sigma }}_\bot} = (\sigma_x, \sigma_y)$.
Associated with the vector potential is the flux density ${n_\phi } = (\nabla  \times  {\bf{A}})_z / {{2\pi \hbar}}$,
which shares the same spatial periodicity as ${{\mathbf{B}}_{{\text{eff}}}}$ and can be considered in general as a type of
flux lattice~\cite{Cooper2011}.

\begin{figure}[btp]
  \includegraphics[width=\columnwidth]{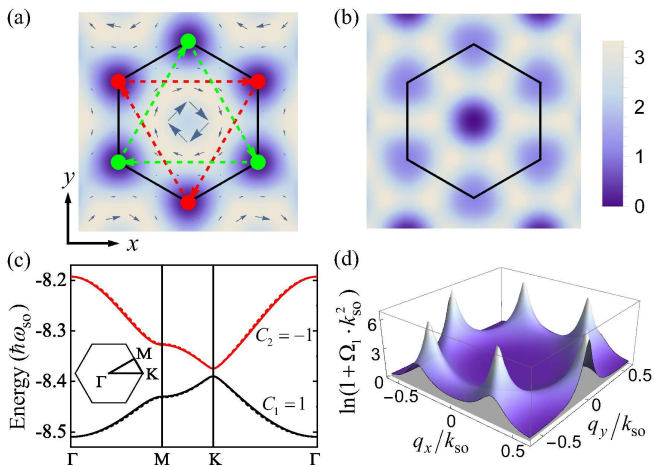}\\
  \caption{Mapping the effective Hamiltonian [Eq.~(\ref{H_eff})] to the Haldane model.
  (a) The density plot of the adiabatic potential $\epsilon_0$ (in blue) and its associated
  vector potential $\mathbf{A}$ shown by field of arrows. 
  A darker color denotes a smaller $\epsilon_0$, the minima of which form a honeycomb lattice
  with two sites per unit cell denoted, respectively, by red and green filled circles.
  The dashed lines denote NNN hopping paths along the directions of positive Peierls phases.
  (b) The flux density $n_\phi$, in units of $10^{-2}k_{\text{so}}^2$.
  The hexagons in (a) and (b) denote the primitive unit cell [the same as in Fig.~\ref{Fig_1}(c)],
  over which the net flux is unity.
  (c) Energy spectrum.
  Solid lines represent the lowest two energy bands along lines with high symmetry in the first Brillouin zone (inserted hexagon) for the effective Hamiltonian Eq.~(\ref{H_eff}), with $\omega_0 = 32.3\omega_{\text{so}}$.
  Dashed lines represent the fitted band structure using the Haldane model results.
  The edge length of the inserted hexagon is $ k_{\text{so}}/{\sqrt{3}}$.
  (d) A logarithmic plot of the Berry curvature for the lowest band $\Omega_1$. The integration of $\Omega_1$ in the first Brillouin zone gives its Chern number, $C_1 = 1$.
  }\label{Fig_2}
\end{figure}

The {{adiabatic potential}} $\epsilon_0$, vector potential $\bf{A}$, and the flux density $n_\phi$ for
our magnetic lattice are shown in Figs.~\ref{Fig_2}(a) and \ref{Fig_2}(b)~\cite{VectorPotential2015}.
As shown in Fig.~\ref{Fig_2}(a), the local minima of $\epsilon_0$ are located at the corners of the unit cell,
forming a honeycomb lattice. When an atom hops between these honeycomb sites, the vector potential contributes
a complex Peierls phase factor $\exp \left( {i\int {{\mathbf{A}}\cdot d{\mathbf{l}}/\hbar} } \right)$ to the hopping coefficient~\cite{Hofstadter1976, Haldane1988, Shao2008},
with the integration evaluated along the corresponding hopping path.
As $\bf{A}$ vanishes along the edges of the hexagon, the nearest-neighboring (NN) phase factor is a trivial unity.
While along the NNN hopping paths [dashed lines in Fig.~\ref{Fig_2}(a)], the accumulated phases
are always nonzero.
Thus the adiabatic scalar potential together with the geometric vector potential realizes the Haldane model in the tight-binding limit.
As a caveat, our flux pattern shown in Fig.~\ref{Fig_2}(b) is not the same as that
suggested by Haldane \cite{Haldane1988}, where the staggered flux density gives a
vanishing net flux over a unit cell.
The flux distribution shown in Fig.~\ref{Fig_2}(b) is non-negative everywhere,
and the net flux over one unit cell is unity rather than zero; this can be checked by integrating the following over a unit cell:
${N_{\phi}} = \frac{1}{4\pi}\int_{{\text{UC}}} {dx\,dy\,\left( {{\mathbf{m}} \cdot {\partial _x}{\mathbf{m}} \times {\partial _y}{\mathbf{m}} } \right)}$, with ${\mathbf{m}} = {{\mathbf{B}}_{{\text{eff}}}}/|{{\mathbf{B}}_{{\text{eff}}}}|$~\cite{Cooper2011}. Thus, the nontrivial winding pattern of 
$ \mathbf{B}_{\rm eff} $ shown in Fig.~\ref{Fig_1}(c) leads to a quantized net flux ${N_\phi } = 1$.
A nonzero net flux generally leads to larger Peierls phases (of order unity).
It also facilitates simulation of charged particles in strong magnetic field with nondispersive Landau levels~\cite{Cooper2011, Cooper2011b}.

To quantitatively confirm that our model indeed maps onto the Haldane model,
we numerically study the spectrum and Berry curvature~\cite{Xiao2011} of the effective Hamiltonian Eq.~(\ref{H_eff})
using the plane-wave expansion method~\cite{Ashcroft1976, SuppMat2015}.
The typical band structure
and the Berry curvature for the lowest band are shown, respectively, in Figs.~\ref{Fig_2}(c) and \ref{Fig_2}(d).
A band gap opens at the corners of the first Brillouin zone ($\pm K$ points),
where the Berry curvature is at a maximum. Both the eigenenergies and the Berry curvatures
are even functions of quasimomentum, so the spectrum at $K'=-K$ is not shown.
The Chern numbers~\cite{TKNN1982} for the lowest two bands are $C_{1,2} = \pm 1$, respectively.
The spectrum and the Berry curvature thus resemble the ones from the Haldane model.
To further validate this correspondence, we adopt the method {{used in}} Ref.~\cite{Ibanez-Azpiroz2014}
to get the NN hopping constant $t_1$ and the complex NNN hopping constant
$|t_2| e^{i\phi}$ of the Haldane model from the calculated band structure.
We find
$t_1 = 0.053\hbar\omega_{\rm so}$ and $|t_2| = 0.0037\hbar\omega_{\rm so}$ with $\phi = 0.40$.
Using these three parameters together with an overall energy shift,
the tight-binding band structure of the Haldane model is plotted with dashed lines in Fig.~\ref{Fig_2}(c).

\begin{figure}[t]
  \includegraphics[width=\columnwidth]{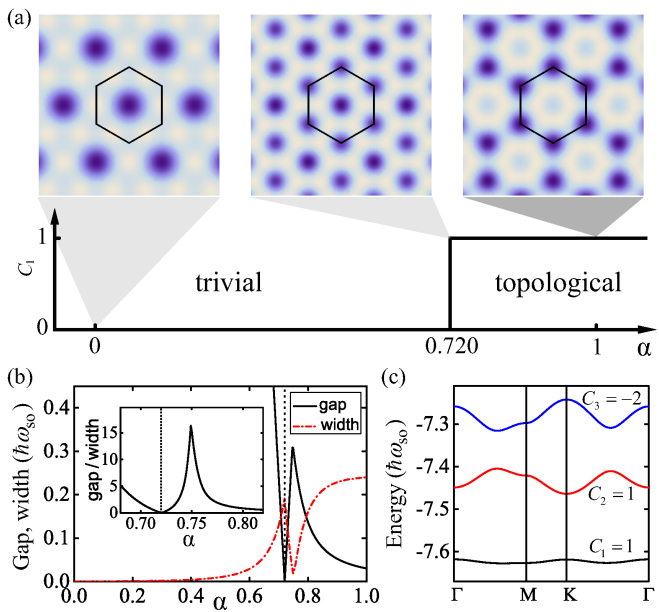}
  \caption{Illustration of the structural and topological phase transition with the tuning of the $z$ component of ${\bf B}_{\rm eff}$ in Eq.~(\ref{B_eff_z_new}). (a) Top panel: the density plots of the
  adiabatic potentials for $\alpha = 0$ (left), $\alpha = 0.720$ (middle), and $\alpha = 1$ (right)
   with the minima (dark colors) forming a simple triangular lattice, a decorated triangular lattice, and a honeycomb lattice, respectively.
   Bottom panel: The Chern number for the lowest energy band as a function of $\alpha$.
   A topological phase transition occurs at the critical point of $\alpha = 0.720$.
  (b) The band gap between the lowest two bands and the band width for the lowest band, as a function of $\alpha$. The band gap closes at $\alpha = 0.720$ (marked by the dotted vertical line). The inset shows the dependence of the band gap-over-width ratio on $\alpha$. It peaks at $\alpha = 0.749$ with a value of $16$.
  (c) The lowest three bands at $\alpha = 0.749$.
  } \label{Fig_3}
\end{figure}

\emph{Topological phase transition and quasiflat bands.}---Our protocol allows for the
easy tuning of two parameters: the SOC strength ${k_{\text{so}}} = \delta t'{g_F}{\mu _B}B'/\hbar $  and
the bias magnetic field $B_0$. Both can be tuned continuously, and can be
turned on adiabatically to reach the ground state for our
model system Eq.~(\ref{H_eff}) \cite{Goldman2014, Jotzu2014} (see~\cite{SuppMat2015} for details).
Once the ground state is achieved, we can apply an additional weak optical gradient field (which commutes with all the pulse manipulation operations) in the $x$-$y$ plane to drive Bloch oscillations and then measure the perpendicular center-of-mass drift to extract the topological properties for the lowest energy band~\cite{Dauphin2013, Jotzu2014, Aidelsburger2015}.
With unequal durations between subsequent pulse pairs,
or allowing for specific $k_{\text{so}}$ and $B_0$ values
for different subperiods, several variants of the effective Hamiltonian
can be synthesized.
A topological phase transition for the lowest energy band can be achieved by a simple tuning of the bias magnetic field.
For this to occur, we set the field strength to be $B_0$ for the first three subperiods
and switch to $\alpha B_0$ for the fourth subperiod;
our protocol then leads to a change for the $z$ component
of the effective magnetic field in Eq.~(\ref{B_eff}) as
\begin{eqnarray}
\label{B_eff_z_new}
{B_{{\text{eff}},z}} = \frac{{{B_0}}}{4}\left[ { - \alpha  + \sum\nolimits_j {\cos \left({k_{{\text{so}}}}{r_{{\theta _j}}}\right)} } \right].
\end{eqnarray}
The $\alpha = 1$ case corresponds to the original proposal with topological bands,
while the $\alpha =0$ case describes a system of trivial energy bands with zero Chern numbers.
By continuously tuning $\alpha$ from $0$ to $1$, a topological phase transition
with band touching and reopening takes place, as summarized in Fig.~\ref{Fig_3}.

Figure~\ref{Fig_3}(a) presents the changing Chern number and, hence, the band topology,
for the lowest band with increasing $\alpha$. The lattice geometry of
the adiabatic potential is found to undergo a structural transformation
first from a simple triangular lattice to a decorated triangular lattice~\cite{Jo-Kurn2012}, and then to a honeycomb lattice.
The corresponding tight-binding descriptions for $s$ orbitals
involve $1$, $3$, and $2$ bands, respectively, for the three cases.
As $\alpha$ increases, the band originating from hopping between
$s$ orbitals located at unit cell centers crosses the two bands
from $s$ orbitals located at the corners. Their corresponding Chern numbers
change after band touching and reopening.
Figure~\ref{Fig_3}(b) shows the behavior of the gap between the lowest two bands
as well as the band width for the lowest one.
Gap closing occurs at the $\Gamma$ {{point}} when $\alpha = 0.720$,
and the gap-over-width ratio is found to be quite large over a limited range after a gap opening
with a peak value as large as $16$ when $\alpha = 0.749$, as shown in the inset of Fig.~\ref{Fig_3}(b).
The energy spectrum at $\alpha = 0.749$ is shown in Fig.~\ref{Fig_3}(c).
The lowest band is a Landau-level-like topological quasiflat one~\cite{Sorensen2005}.
Such a nondispersive topological band also persists beyond the adiabatic limit~\cite{SuppMat2015}.
It is a promising candidate for simulating the fractional quantum Hall effect
when suitable interactions are included~\cite{Yang-Gu2012, Parameswaran2013, Cooper2013, Bergholtz-Liu2013}.
It is perhaps worth pointing out, also, that a flat band can emerge as the second excited band in a Kagome lattice~\cite{Jo-Kurn2012}, or as the first excited band in a Lieb lattice~\cite{Taie2015}. The properties of the localized states in the flat band of a Lieb lattice have been investigated in a recent experiment~\cite{Taie2015}.

We focus in this Letter on discussing single-particle physics of a fermionic spin-$1/2$ system, though our magnetic lattice generation protocol can be
equally applied to a higher-spin atom, be it a boson or fermion. When local momentum-independent ($s$ wave) interaction is taken into account, it can be simply added to the effective Hamiltonian because it commutes with all the pulse manipulation operations (see also~\cite{Xu2013, Anderson2013}). The topological phase is expected to be stable for weak interactions due to the presence of a gap. However, stronger interaction can drive the system to new phases, in which the physics may be dominated by the interplay between the correlation and band topology. A detailed study of the interaction effects for this system deserves further efforts.

In conclusion, we propose an experimentally feasible protocol
to realize a synthetic magnetic field with real magnetic field pulses. The synthetic magnetic field forms a lattice with nontrivial band topology, and under certain limits can be mapped to the Haldane model.
The high tunability of our scheme makes it possible to design a topological phase transition as well as
quasiflat energy bands with nontrivial topology, which could push
the effective model into the strongly correlated regime.

We thank Professors W.~Vincent Liu and Kun Yang for valuable discussions. This work is supported by the MOST Grant No. 2013CB922004 of the National Key Basic Research Program of China and by the NSFC (Grants No.~91121005, No.~91421305, and No.~11374176), as well as by U.S. AFOSR (Grant No.~FA9550-12-1-0079), ARO (Grant No.~W911NF-11-1-0230), the Charles E. Kaufman Foundation, and the Pittsburgh Foundation (J.Y. and Z.-F. X.). R.L. also wants to acknowledge support from the NSFC (Grant No. 11274195).

\newpage
\onecolumngrid
\renewcommand\thefigure{S\arabic{figure}}
\setcounter{figure}{0}
\renewcommand\theequation{S\arabic{equation}}
\setcounter{equation}{0}
\newpage
{
\center \bf \large
Supplemental Material for: \\
Dynamical Generation of Topological Magnetic Lattices for Ultracold Atoms\vspace*{0.1cm}\\
\vspace*{0.0cm}
}
\begin{center}
Jinlong Yu$^{1,3}$, Zhi-Fang Xu$^{2,3}$, Rong L\"u$^{1,4}$, and Li You$^{1,4}$\\
\vspace*{0.15cm}
\small{\textit{$^1$State Key Laboratory of Low Dimensional Quantum Physics,\\ Department of Physics, Tsinghua University, Beijing 100084, China\\
$^2$MOE Key Laboratory of Fundamental Physical Quantities Measurements, \\ School of Physics, Huazhong University of Science and Technology, Wuhan 430074, China\\
$^3$Department of Physics and Astronomy, University of Pittsburgh, Pittsburgh, Pennsylvania 15260, USA\\
$^4$Collaborative Innovation Center of Quantum Matter, Beijing 100084, China}}\\
\vspace*{0.25cm}
\end{center}

This supplementary material provides more details
for the various points mentioned in the main article.

\subsection{Derivation of the effective Hamiltonian}

As explained in the main text, a gradient magnetic field pulse along $\hat e_\theta = \left(\cos\theta, \sin\theta, 0\right)$ in the $x$-$y$ plane prints a phase factor $\exp \left( -{i{k_{\rm so}}r_\theta{F_\theta}} \right)$ onto the single particle wave function.
Under the action of an opposite pulse pair, the atomic momentum {{$\mathbf{p}=\hbar\mathbf{k}$}} and spin $F_z$ transform according to,
\begin{eqnarray}
  \begin{aligned}
  {e^{  i{k_{\rm so}}{r_\theta }{F_\theta }}}{k_x}{e^{-i{k_{\rm so}}{r_\theta }{F_\theta }}} &= {k_x} - {k_{\rm so}}{F_\theta }\cos \theta,  \hfill \\
  {e^{  i{k_{\rm so}}{r_\theta }{F_\theta }}}{k_y}{e^{-i{k_{\rm so}}{r_\theta }{F_\theta }}} &= {k_y} - {k_{\rm so}}{F_\theta }\sin \theta,  \hfill \\
  {e^{  i{k_{\rm so}}{r_\theta }{F_\theta }}}{F_z}{e^{-i{k_{\rm so}}{r_\theta }{F_\theta }}} &= {F_z}\cos \left( {{k_{\rm so}}{r_\theta }} \right) - {{\tilde F}_\theta }\sin \left( {{k_{\rm so}}{r_\theta }} \right), \hfill \\
\end{aligned}
\end{eqnarray}
where ${{\tilde F}_\theta } \equiv {F_x}\sin \theta  - {F_y}\cos \theta $ is the spin operator perpendicular to $\hat e_\theta$.
Thus the evolution including the pulse pair gives rise to
\begin{eqnarray} \label{U_theta}
  {U_\theta }\left( {\delta t,0} \right) = {e^{  i{k_{\rm so}}{r_\theta }{F_\theta }}}{e^{ - i{H_0}{{\delta t}}/{\hbar }}}{e^{-i{k_{\rm so}}{r_\theta }{F_\theta }}} \equiv {e^{ - i{H_{\theta }} {{\delta t}}/{\hbar } }},
\end{eqnarray}
with the corresponding Hamiltonian
\begin{eqnarray} \label{Hami_along_theta}
  {H_{\theta }}  &= &  {e^{  i{k_{\rm so}}{r_\theta }{F_\theta }}}{H_0}{e^{-i{k_{\rm so}}{r_\theta }{F_\theta }}} \nonumber \\
&=& \frac{{{\hbar ^2}}}{{2m}}{\left( {{k_x} - {k_{\rm so}}{F_\theta }\cos \theta } \right)^2} + \frac{{{\hbar ^2}}}{{2m}}{\left( {{k_y} - {k_{\rm so}}{F_\theta }\sin \theta } \right)^2}  \nonumber \\
  & &+ \hbar {\omega _0}\left[ {{F_z}\cos \left( {{k_{\rm so}}{r_\theta }} \right) - {{\tilde F}_\theta }\sin \left( {{k_{\rm so}}{r_\theta }} \right)} \right].
\end{eqnarray}

A {{$\mp\pi$}} pulse pair along the $y$ direction flips the direction of the magnetic field seen by atom
without inducing coupling to momentum. The evolution operator from such a pulse pair is given by
\begin{eqnarray} \label{U_theta}
  {U_{y, \pi} }\left( {\delta t,0} \right) = {e^{  -i \pi {F_y }}}{e^{ - i {H_0}{{\delta t}}/{\hbar }}}{e^{i \pi{F_y }}} \equiv {e^{ - i{H_{y , \pi}} {{\delta t}}/{\hbar } }},
\end{eqnarray}
with the corresponding Hamiltonian
\begin{eqnarray}\label{H_pi_pulse}
  {H_{y , \pi}} = \frac{\hbar^2 }{2m} \left( {{k}}^2_x + {{k}}^2_y\right) - \hbar \omega_0 F_z.
\end{eqnarray}

A complete evolution period contains three gradient pulse pairs
separated by a mutual angle of {{$120^\circ$}} in the $x$-$y$ plane,
together with a $\mp\pi$ pulse pair along the $y$ direction,
the corresponding evolution operator for one period $T=4\delta t$ is given by
\begin{eqnarray}\label{U_eff}
 U\left( {T = 4\delta t,0} \right) &=& {e^{ - i{H_{y,\pi }}\delta t/\hbar }}{e^{ - i{H_{{\theta _3}}}\delta t/\hbar }}{e^{ - i{H_{{\theta _2}}}\delta t/\hbar }}{e^{ - i{H_{{\theta _1}}}\delta t/\hbar }}\nonumber\\
 & \equiv & {e^{ - i{H_{{\text{eff}}}}T/\hbar }},
\end{eqnarray}
with ${\theta _j = -\frac{\pi}{6} + \frac{2\pi j}{3} }$.
We use $H_m$ to denote the Hamiltonian for each subperiods as
\begin{equation}
  {H_1} = {H_{{\theta _1}}},\;{H_2} = {H_{{\theta _2}}},\;{H_3} = {H_{{\theta _3}}},\;{H_4} = {H_{y,\pi }}.
\end{equation}
To lowest order of the Trotter's expansion for $T$, the effective Hamiltonian ${H_{\rm eff}}$ is the average of $H_m$,
\begin{eqnarray}\label{H_eff_add}
{H_{{\text{eff}}}}  = \frac{1}{N}\sum\limits_{m = 1}^N {{H_m}},
\end{eqnarray}
with $N=4$ for the current case.
Expanding out the terms in this Hamiltonian explicitly, we arrive
at the effective Hamiltonian as shown in the main text after recombining them.

The higher order corrections of the effective Hamiltonian can also be derived  through the Magnus expansion method, which to first order of $T$ gives~\cite{Slichter1990}
\begin{equation}\label{H_eff_1_order}
{H_{{\text{eff}}}} = \frac{1}{N}\sum\limits_{m = 1}^N {{H_m}}  + \frac{{iT}}{{2\hbar {N^2}}}\sum\limits_{m < n = 2}^N {[{H_m},{H_n}]}.
\end{equation}

From Eq.~(\ref{U_eff}) we see that, our protocol can also be viewed as a periodically driven 4-step sequence, where the dynamics of the driven system can be understood in terms of the effective Hamiltonian and the associated micromotion~\cite{Goldman2014}. We now provide the details for how our system fits into this formulism.
The Hamiltonian for a general N-step sequence is described by ${H_m} = {H^{(0)}} + {V_m}$,
with the constraint $\sum\nolimits_{m = 1}^N {{V_m}}  = 0$. Thus we find
\begin{equation}\label{H_aver}
  {H^{(0)}} = \frac{1}{N}\sum\limits_{m = 1}^N {{H_m}},
\end{equation}
which is just the time-averaged effective Hamiltonian Eq.~(\ref{H_eff_add}), and correspondingly
\begin{equation}\label{V_m}
  {V_m} = {H_m} - \frac{1}{N}\sum\limits_{m = 1}^N {{H_m}}.
\end{equation}
The evolution operator for such a driven system can be partitioned in the following as introduced in Ref.~\cite{Goldman2014}
\begin{equation}
  U\left( {t,0} \right) = {e^{ - iK(t)}}{e^{ - it{{\tilde H}_{{\text{eff}}}}/\hbar }}{e^{iK(0)}},
\end{equation}
where the effective Hamiltonian ${\tilde H}_{\text{eff}}$ and the initial-kick operator $K(0)$ of the N-step sequence to first order of $T$ are given by~\cite{Goldman2014}
\begin{equation}\label{H_Kick}
  \begin{aligned}
  {{\tilde H}_{{\text{eff}}}} & = {H^{(0)}}+{H^{(1)}} = {H^{(0)}} + \frac{{ i T}}{{{N^3} }}\sum\limits_{m < n = 2}^N {{\mathcal{C}_{m,n}}[{V_m},{V_n}]} , \hfill \\
  K\left( 0 \right) &= \frac{{T }}{{{N^2}\hbar  }}\sum\limits_{m = 1}^N {{V_m}m} , \hfill \\
  \end{aligned}
\end{equation}
with ${\mathcal{C}_{m,n}} = \frac{N}{2} + m - n$.
For a full evolution cycle ($t = T$), the evolution operator is found to be
\begin{equation}
  U\left( {T,0} \right) = {e^{ - iK(T)}}{e^{ - iT{{\tilde H}_{{\text{eff}}}}/\hbar }}{e^{iK(0)}} = {e^{ - iT{H_{{\text{eff}}}}/\hbar }},
\end{equation}
where the Hamiltonian ${{H_{{\text{eff}}}}}$ to first order of $T$ is given by [using $K(T)=K(0)$]
\begin{equation}
  {H_{{\text{eff}}}} = {e^{ - iK(T)}}{{\tilde H}_{{\text{eff}}}}{e^{iK(0)}} \simeq {H^{(0)}} + {H^{(1)}} - i[K(0),{H^{(0)}}].
\end{equation}
Inserting Eqs.~(\ref{H_aver}), (\ref{V_m}) and (\ref{H_Kick}) into the above equation, we recover Eq.~(\ref{H_eff_1_order}) as expected.

This shows that, the micromotion of the N-step periodically driven system gives the first order correction to the effective Hamiltonian. This calls for the use of the time-averaged effective Hamiltonian Eq.~(\ref{H_eff_add}) to approximate the evolution in the short $T$ limit. This approximation will be validated numerically in the later sections.

\subsection{Solving the effective Hamiltonian with {{the plane wave}} expansion method}

For spin-1/2 with $\mathbf{ F} =  \bm{\sigma}/{2}$, the effective Hamiltonian takes the following form
\begin{eqnarray}\label{H_half_eff}
  {H_{{\rm{eff}}}} = \frac{1}{{2m}}{\left( {{\mathbf{p}} - \frac{3}{{16}}\hbar {k_{{\text{so}}}}{{\bm{\sigma }}_\bot} } \right)^2} + \frac{1}{2}{g_F}{\mu _B}{{\mathbf{B}}_{{\text{eff}}}}\cdot{\bm{\sigma }},
\end{eqnarray}
with ${{\bm{\sigma }}_\bot} = (\sigma_x, \sigma_y)$ after omitting a constant energy shift $15\hbar {\omega _{{\text{so}}}}/128$.
As $H_{\rm eff}$ in  Eq.~(\ref{H_half_eff}) is space-periodic, its
eigenstates can be labeled with
quasimomentum $\mathbf{q} = \left(q_x, q_y\right)$ in the first Brillouin zone as good quantum numbers.
Expanded in the plane wave basis \cite{Ashcroft1976}, the eigenstates take the form
\begin{eqnarray}\label{plane_wave_expansion_psi}
  {\psi _{n\mathbf{q}}}\left( {\mathbf{r}} \right) = {e^{i\mathbf{q} \cdot \mathbf{r}}}\sum\limits_{l,m} {{e^{i\left( {l{{
  \mathbf{b}}_1} + m{{\mathbf{b}}_2}} \right) \cdot \mathbf{r}}}} \left( \begin{gathered}
  {C_{l,m, \uparrow }} \hfill \\
  {C_{l,m, \downarrow }} \hfill \\
\end{gathered}  \right),
\end{eqnarray}
where $n$ is the band index, ${{\mathbf{b}}_{1,2}} = {k_{\rm so}}\left({ \mp \cos \left( {\frac{\pi }{6}} \right), - \sin \left( {\frac{\pi }{6}} \right)} \right)$ are the two reciprocal unit vectors, $l, m$ are integers taking values $0, \pm 1, \pm 2,$ etc.
The expansion coefficients $C_{l,m,\uparrow(\downarrow)}$ are to be determined.
The plane wave expansion Eq.~(\ref{plane_wave_expansion_psi}) together with the eigenvalue equation
\begin{eqnarray}
{H_{\rm eff}}{\psi _{n\mathbf{q}}}\left( {\mathbf{r}} \right) = E_n\left( {\mathbf{q}} \right){\psi _{n\mathbf{q}}}\left( {\mathbf{r}} \right),
\end{eqnarray}
determines the energy spectrum, as well as the Bloch wave functions as eigenfunctions.

In our numerical calculations, we use a cutoff $N = 15$ as the maximal value for $|l|$ and $|m|$. And we have checked that, for $N>10$ cases, the band structures shown in the main text are independent of the choice of $N$. Thus in the plane wave basis, the Hamiltonian is expressed as a $[2\left( {2N + 1} \right)^2] \times [2\left( {2N + 1} \right)^2]$ matrix for each quasimomentum $\mathbf{q}$ within the first Brillouin zone. The eigenvalues and corresponding eigenvectors give respectively the energy spectrum and Bloch wave functions.

Once the Bloch wave functions are obtained, we can calculate their Berry curvatures ${\Omega _n}\left( {\mathbf{q}} \right)$ and (first) Chern numbers ${C_n}$ for each energy bands according to \cite{Xiao2011}
\begin{eqnarray}
   \begin{aligned}
  {\Omega _n}\left( {\mathbf{q}} \right) &= i{\left[ {{\nabla _{\mathbf{q}}} \times \left\langle {{u_n}(\mathbf{q})} \right|{\nabla _{\mathbf{q}}}\left| {{u_n}(\mathbf{q})} \right\rangle } \right]_z}, \hfill \\
  {C_n} &= \frac{1}{{2\pi }}\int_{{\text{BZ}}} {{d^2}q{\Omega _n}\left( {\mathbf{q}} \right)} , \hfill \\
\end{aligned}
\end{eqnarray}
where ${u_{n\mathbf{q}}}(\mathbf{r}) = \left\langle {\mathbf{r}\left| {{u_n}(\mathbf{q})} \right.} \right\rangle  = {e^{ - i\mathbf{q} \cdot \mathbf{r}}}{\psi _{n\mathbf{q}}}(\mathbf{r})$ is the cell-periodic part of the Bloch function.

\subsection{Validity of the effective Hamiltonian}

We use the effective Hamiltonian to describe the evolution of ultracold atoms under the magnetic field pulse sequence
of the proposed protocol. The use of Trotter expansion limits its validity to short evolution periods.
In this subsection we use two complimentary methods to check for this approximation.

 \begin{figure}[hbtp]
  \includegraphics[width=0.5\columnwidth]{{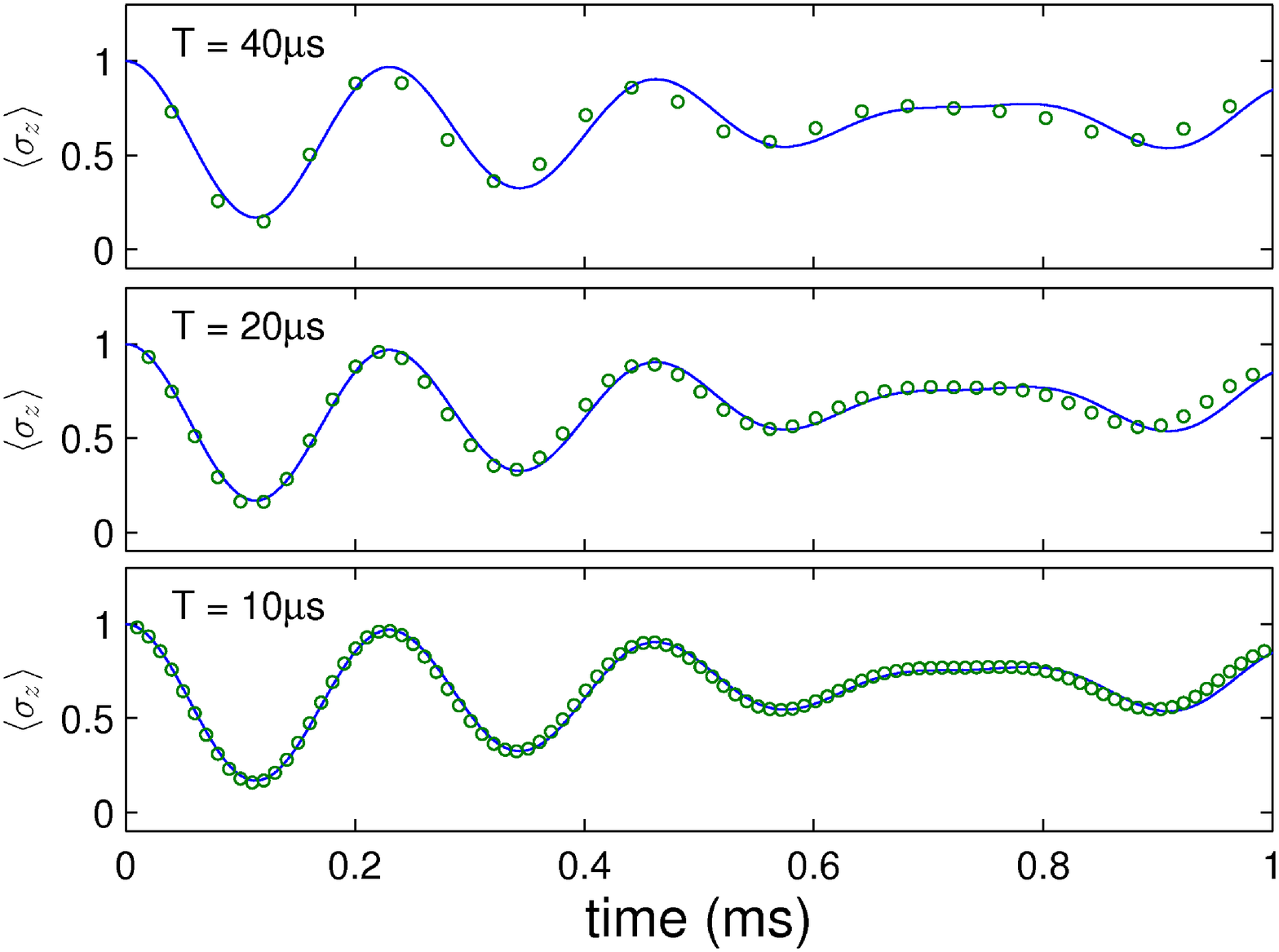}}\\
  \caption{The evolution of a Gaussian wave packet located at the center of a harmonic trap, with oscillator frequencies $\omega_x = \omega_y = (2\pi)\times 30\,\text{Hz}$ is investigated.  The system is initially prepared in the spin up state. The figure shows the time dependent population imbalance $\left\langle {{\sigma _z}} \right\rangle$ calculated from the effective Hamiltonian (solid line), and the actual pulse sequence (open circles), with evolution periods $T = 40\,\mu \rm{s}$, $20\,\mu \rm{s}$, and $10\,\mu \rm{s}$, from top to down respectively.
  The other parameters are $\omega_0 = 32.3\omega_{\rm so}=(2\pi)\times9.3\,\text{kHz}$, $\alpha = 1$.
  }\label{Fig_S1}
 \end{figure}

The first method replies on evolving an eigenstate, whereby the initial state ${\psi _{n\mathbf{q}}}\left( {\mathbf{r}}, t=0 \right)$
is prepared in an eigenstate of the effective Hamiltonian with a given band index $n$ and quasimomentum $\mathbf{q}$.
The actual magnetic field pulse sequence is then used to evolve it to a later time $t >0$ where the
wave function is denoted by ${\phi _{n\mathbf{q}}}\left( {\mathbf{r}}, t \right)$. The overlap between these two states $ \langle{\phi_{n\mathbf{q}}\left(t\right)| \psi_{n\mathbf{q}}\left(t=0\right)\rangle}$
should be identically equal to {\textcolor{black}{unity}} at all times for the ideal case when
the effective Hamiltonian exactly represents the evolution by the pulse sequence.

We arbitrarily choose a quasimomentum point $\mathbf{q} = (0.3, -0.2){{k_{\text{so}}}}$ in the first Brillouin zone,
and the eigenstate for the lowest band $n = 1$, for the case $\omega_0 = 32.3\omega_{\rm so}=(2\pi)\times9.3\,\text{kHz}$ and $\alpha = 1$, is calculated according to the method described in the last section.
The absolute values of the overlap after evolving for $1\,\rm{ms}$ for $T = 40\,\mu \rm{s}$, $20\,\mu \rm{s}$ and $10\,\mu \rm{s}$,
are respectively calculated to be $0.947$, $0.984$ and $0.995$, indicating improved level of approximations with
shorter periods.

The second method relies on wave packet evolution. For this case,
an external trapping potential ${V_{{\text{trap}}}}(x,y) = \frac{1}{2}m(\omega _x^2{x^2} + \omega _y^2{y^2})$
is included into the Hamiltonian, and the initial state is prepared as a wave packet in the harmonic trap.
Since the trap potential commutes with all the pulses, it is simply added into the effective Hamiltonian.
The initial wave packet is then evolved respectively through the effective Hamiltonian and
through the actually pulse sequence. We can then calculate some physical observables, e.g.,
population imbalance, by the corresponding time-dependent states, and compare
their respective results. Figure~\ref{Fig_S1} shows an example of this comparison.
We can see that, they coincide with each other better for shorter evolution periods.

These studies indicate that for sufficiently short evolution periods,
the effective Hamiltonian faithfully describes the evolution under the actual magnetic field pulse sequence.
For the cases we discuss, $T = 10 \,\mu \rm{s} \lesssim 1/\omega_0 = 17 \,\mu \rm{s}$
is short enough to validate the effective Hamiltonian approximation.

\subsection{Derivation of the emergent gauge fields}

The evolution of a single-particle state is governed by the effective Hamiltonian according to
\begin{eqnarray}\label{Schro_original}
  i\hbar \frac{\partial }{{\partial t}}\left| {\Psi \left( {{\mathbf{r}},{\mathbf{t}}} \right)} \right\rangle  = {H_{{\text{eff}}}}\left| {\Psi \left( {{\mathbf{r}},{\mathbf{t}}} \right)} \right\rangle ,
\end{eqnarray}
with ${H_{{\rm{eff}}}}$ given by Eq.~(\ref{H_half_eff}) for the spin $1/2$ case.
Expanded in the adiabatic basis, with corresponding center-of-mass wave functions
labeled as ${{\psi _j}\left( {{\mathbf{r}},{\mathbf{t}}} \right)}$, the wave function takes the form
\begin{eqnarray}
  \left| {\Psi \left( {{\mathbf{r}},{\mathbf{t}}} \right)} \right\rangle  = \sum\limits_{j = 1,2} {{\psi _j}\left( {{\mathbf{r}},{\mathbf{t}}} \right)} \left| {{\chi _j}\left( {\mathbf{r}} \right)} \right\rangle,
\end{eqnarray}
which when acted upon by the spin-dependent shift leads to
\begin{eqnarray}
  \left( {{\mathbf{p}} - \frac{3}{{16}}\hbar {k_{{\text{so}}}}{ {\bm{ \sigma }}_\bot}} \right)\left| {\Psi \left( {{\mathbf{r}},{\mathbf{t}}} \right)} \right\rangle  = \sum\limits_{l,j = 1}^2 {\left( {{\mathbf{p}}{\delta _{l,j}} - {{\mathbf{A}}_{lj}} - {{{\mathbf{\tilde A}}}_{lj}}} \right){\psi _j}} \left| {{\chi _l}} \right\rangle, \hskip 24pt
\end{eqnarray}
with
\begin{eqnarray}\label{A_ij_tilde}
  \begin{gathered}
  {{\mathbf{A}}_{lj}} = i\hbar \left\langle {{\chi _l}} \right|\left. {\nabla {\chi _j}} \right\rangle , \hfill \\
  {{{\mathbf{\tilde A}}}_{lj}} = \frac{3}{{16}}\hbar {k_{{\text{so}}}}\left\langle {{\chi _l}} \right|{ {\bm{ \sigma }}_\bot}\left| {{\chi _j}} \right\rangle . \hfill \\
\end{gathered}
\end{eqnarray}

In the adiabatic limit $\omega_0\gg\omega_{\rm so}$, if the initial state is prepared in the dressed state $\left| {{\chi _1}} \right\rangle$, then the probability amplitude for the particle to be in the orthogonal state $\left| {{\chi _2}} \right\rangle$ remains zero at all time. Thus by projecting Eq.~(\ref{Schro_original}) to the dressed state $\left| {{\chi _1}} \right\rangle$ and taking $\psi_2 = 0$, we get a closed equation for $\psi_1$
\begin{eqnarray}\label{Schro_psi1}
  i\hbar \frac{\partial }{{\partial t}}{\psi _1} = \left[ {\frac{1}{{2m}}{{\left( {{\mathbf{p}} - {\mathbf{A}}} \right)}^2} + {\epsilon _0} + W} \right]{\psi _1},
\end{eqnarray}
where
\begin{eqnarray}
  {\mathbf{A}} \equiv {{\mathbf{A}}_{11}} + {{{\mathbf{\tilde A}}}_{11}} = i\hbar \left\langle {{\chi _1}} \right|\left. {\nabla {\chi _1}} \right\rangle  + \frac{3}{{16}}\hbar {k_{{\text{so}}}}\left\langle {{\chi _1}} \right|{{\bm{\sigma }}_\bot}\left| {{\chi _1}} \right\rangle,
\end{eqnarray}
is the geometric vector potential that couples to the center-of-mass motion, and
\begin{eqnarray}\label{W_potential}
 W = \frac{1}{{2m}}\left( {{{\mathbf{A}}_{12}} \cdot {{\mathbf{A}}_{21}} + {{{\mathbf{\tilde A}}}_{12}} \cdot {{{\mathbf{\tilde A}}}_{21}} + {{\mathbf{A}}_{12}} \cdot {{{\mathbf{\tilde A}}}_{21}} + {{{\mathbf{\tilde A}}}_{12}} \cdot {{\mathbf{A}}_{21}}} \right), \hskip 24pt
\end{eqnarray}
is the geometric scalar potential.

 \begin{figure}[hbtp]
  \includegraphics[width=0.7\columnwidth]{{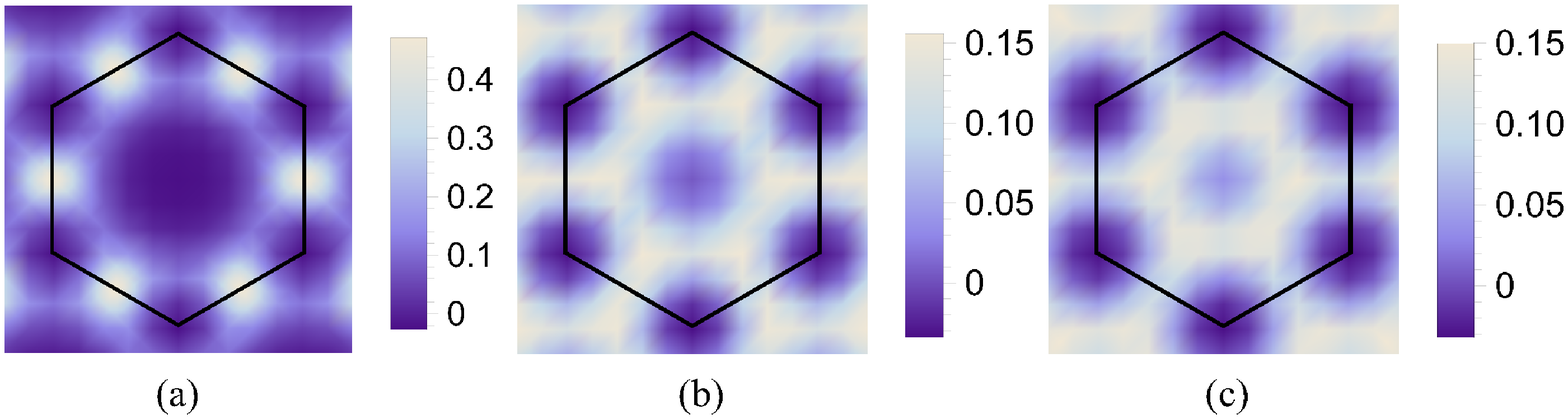}}\\
  \caption{The geometric scalar potentials W (in units of $\hbar\omega_{\text{so}}$) in Eq.~(\ref{W_potential}) at $\omega_{0} = 32.3\omega_{\rm so}$ for (a) $\alpha = 0$, (b) $\alpha = 0.72$, and (c) $\alpha = 1$.   
  }\label{Fig_S2}
 \end{figure}

The resulting scalar potential consists of two terms. One is the adiabatic potential $\epsilon_0$,
whose characteristic energy scale is $\hbar \omega_{ 0}$. The other is the geometric scalar potential $W$,
which scales as $\hbar \omega_{\rm so}$. Thus $W$ makes a negligible contribution to the total scalar potential
when compared to $\epsilon_0$, in the $\omega_0\gg\omega_{\rm so}$ limit.
In Fig.~\ref{Fig_S2}, geometric potentials for $\alpha = 0$, $0.72$, and $1$ at $\omega_{0} = 32.3\omega_{\rm so}$
are shown. Compared to the corresponding adiabatic potentials shown respectively in Fig.~3(a) in the main text, with the corresponding potential (minima, maxima) in units of $\hbar\omega_{\text{so}}$ being $(-12.0, -4.0)$, $(-9.2, -6.7)$ and $(-10.0, -7.0)$ respectively,
the geometric potentials only give small corrections to the final total scalar potentials.
The qualitative understanding of the Haldane model as well as the topological phase transition
enabled by our model is not affected by including these corrections.

If we take ${\tilde {\bf A}}_{lj} = 0$ in Eq.~(\ref{A_ij_tilde}), then the above results
reduce to those reviewed by Dalibard \emph{et al.} in Ref. \cite{Dalibard2011},
where momentum is not coupled to spin components in the original Hamiltonian,
and the corresponding geometric potentials are ${\mathbf{A}} = i\hbar \left\langle {\left. {{\chi _1}} \right|\nabla {\chi _1}} \right\rangle$ and $W = {{\mathbf{A}}_{12}} \cdot {{\mathbf{A}}_{21}}/2m = {\hbar ^2}|\left\langle {{\chi _1}} \right|\left. {\nabla {\chi _2}} \right\rangle {|^2}/2m$.

\subsection{Energy spectrum beyond the limit of adiabatic approximation}
Although the discussion in the main text on the generation and understanding of the topological energy bands is based on models under the adiabatic as well as tight-binding approximation, the nontrivial topology of the energy bands is found to exist beyond these two approximations~\cite{Cooper2011, Cooper2011b}.
As an example, we take $\alpha = 0.749$, and reduce the original $\omega_0$ value by a factor of five to $\omega_0 = \frac{32.3}{5}\omega_{\text{so}} = (2\pi)\times1.9\,\text{kHz}$.
The term of a Zeeman level is used instead of the adiabatic potential when the adiabatic approximation is not satisfied.
At $\omega_0 = 6.46\omega_{\text{so}}$ with $\alpha = 0.749$,
the (minima,maxima) of the two Zeeman levels of the magnetic lattice are respectively given by $(-1.84, -1.34)$ and $(1.34, 1.84)$ in units of $\hbar \omega_{\rm so}$.
The separation between the corresponding two levels is $2.68\hbar \omega_{\rm so}$,
and the lattice depth of the lower Zeeman level is a mere $0.5\hbar \omega_{\rm so}$.
As the energy scale for the spatial uniform term in Eq.~(\ref{H_half_eff}) is of order $\hbar \omega_{\rm so}$,
and SOC in this term can flip the spin, the lattice term does not dominate during the evolution; the adiabatic approximation fails for this case.

\begin{figure}[hbtp]
  \includegraphics[width=\columnwidth]{{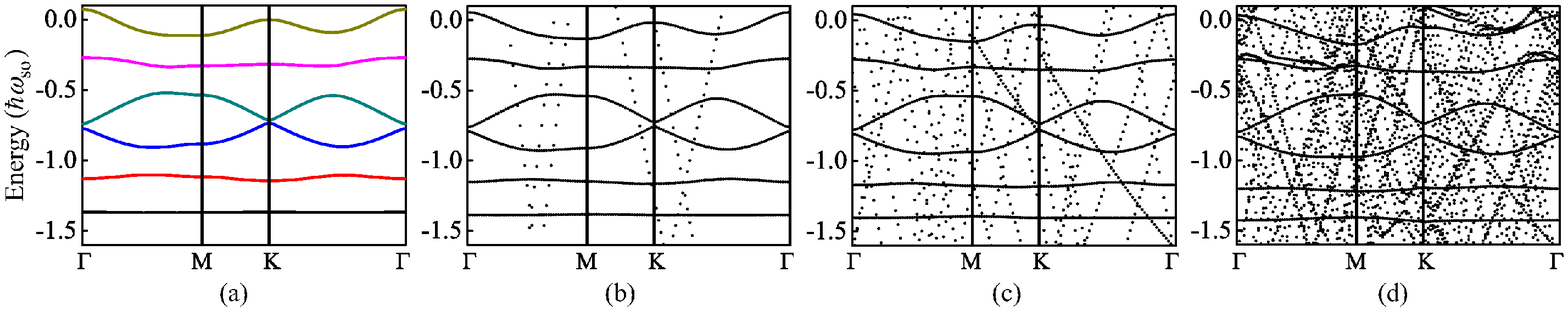}}\\
  \caption{The energy spectrum for our periodically driven model system at $\alpha = 0.749$ and $\omega_0 = 6.46\omega_{\text{so}} = (2\pi)\times1.9\,\text{kHz}$. (a) From the effective Hamiltonian Eq.~(\ref{H_half_eff}). (b-d) The quasienergies from solving the evolution operator in Eq.~(\ref{U_eff}) for (b) $T = 50\,\mu\text{s}$; (c) $T = 100\,\mu\text{s}$; and (d) $T = 200\,\mu\text{s}$.
  }\label{Fig_S3}
 \end{figure}

The lowest six energy bands of the effective Hamiltonian for this case are shown in Fig.~\ref{Fig_S3}(a).
Their Chern numbers are all found to be $C_n = 1$, which invalidates directly a tight-binding description,
because the sum of Chern numbers for a set of tight-binding bands should be equal to zero.
The band width for the lowest energy band is $0.0049\hbar \omega_\text{so}$, and the lowest band gap is $0.22\hbar \omega_{\text{so}}$,
which gives a gap-over-width ratio of $\sim 45$. Thus these Chern numbers for the lowest few bands resemble the ones
for a charged particle in a magnetic field in a weak lattice background~\cite{Ashcroft1976}.

Next we discuss the validity of the effective Hamiltonian for this case. In previous subsection and in Fig.~\ref{Fig_S1},
we show the validity of the effective Hamiltonian for $\omega_0 = 32.3\omega_{\rm so}=(2\pi)\times9.3\,\text{kHz}$,
with respectively evolution periods $T = 10\,\mu \rm{s}$, $20\,\mu \rm{s}$, and $40\,\mu \rm{s}$.
In the present case of $\omega_0 = (2\pi)\times1.9\,\text{kHz}$, i.e., reduced by a factor of five,
the suitable evolution period is expected to be increased by a similar factor.
We can repeat the previous method by scanning state evolution period to illustrate the
validity of adiabatic approximation, or alternatively, we can calculate the Floquet spectrum
of quasienergies by evaluating the evolution operator Eq.~(\ref{U_eff}) directly for different evolution periods,
 with the results as shown in Fig.~\ref{Fig_S3}(b-d). We see that the spectrum of the quasienergies strongly resembles
 the one from the effective Hamiltonian at least for $T = 100\,\mu\rm{s}$.
 The topological properties of the lowest energy band is expected to be preserved for $T = 200\,\mu\rm{s}$,
 because no signature of band closing is found. For even larger evolution period, e.g. $T = 400\,\mu\rm{s}$,
 the quasienergies for the lower energy bands are found to be smeared out due to the wrapping of the higher Floquet
 sectors with quasienergies at multiples of $2\pi\hbar/T$.
 Thus we conclude the suitable evolution period required for the validity of the effective Hamiltonian
approximation is of the order of $\sim 100\,\mu\rm{s}$ for the present choice of $\omega_0 = (2\pi)\times1.9\,\text{kHz}$.

\subsection{Adiabatic preparation of the ground state}
As is verified in the previous sections, one can use the effective Hamiltonian Eq.~(\ref{H_half_eff}) to describe the state evolution under the pulse sequence for short evolution periods. In this section, we describe how can the ground state of the effective Hamiltonian Eq.~(\ref{H_half_eff}) be reached through an adiabatic loading approach. For a non-interacting many-body fermionic system, the ability of preparing its ground state is essentially equivalent to the statement that all the (single-particle) eigenstates in the lowest energy band of ${H_{{\rm{eff}}}}$ in Eq.~(\ref{H_half_eff}) can be reached appropriately.

As shown in the main text, one complete cycle of the evolution period consists of four subperiods. To prepare the ground state, we will keep the fourth subperiod intact, while adiabatically ramp up (a) the strength of the gradient pulses in $N_a$ complete cycles and then (b) the bias magnetic field for the first three subperiods in $N_b$ complete cycles.

A quantitative analysis is provided as follows. We first introduce an auxiliary Hamiltonian
\begin{equation}
  \mathcal{H}\left( {\kappa ,\gamma } \right) = \frac{{{{\mathbf{p}}^2}}}{{2m}} - \frac{{3\hbar {k_{{\text{so}}}}}}{{16m}}\kappa \left( {{p_x}{\sigma _x} + {p_y}{\sigma _y}} \right) - \frac{\alpha }{8}\hbar {\omega _0}{\sigma _z} + \frac{\gamma }{8}\hbar {\omega _0}{\mathbf{M}} \cdot {\bm{\sigma }},
\end{equation}
where the three components of the $\mathbf{M}$ vector are respectively ${M_x} =  - \sum\nolimits_j {\sin \left( {{k_{{\text{so}}}}{r_{{\theta _j}}}} \right)\sin \left( {{\theta _j}} \right)} $, ${M_y} = \sum\nolimits_j {\sin \left( {{k_{{\text{so}}}}{r_{{\theta _j}}}} \right)\cos \left( {{\theta _j}} \right)} $, and ${M_z} = \sum\nolimits_j {\cos \left( {{k_{{\text{so}}}}{r_{{\theta _j}}}} \right)}$. The effective Hamiltonian before the aforementioned ramping process is $\mathcal{H}\left( {\kappa = 0,\gamma = 0 } \right)$; the effective Hamiltonian at the end of the ramping stage (a) is $\mathcal{H}\left( {\kappa = 1,\gamma = 0 } \right)$; and the effective Hamiltonian after ramping stage (b) is $\mathcal{H}\left( {\kappa = 1,\gamma = 1 } \right)$, which recovers the effective Hamiltonian ${H_{{\rm{eff}}}}$ in Eq.~(\ref{H_half_eff}).

As a proof-of-principal illustration, suppose the initial state for the ramping stage (a) is prepared as a plain wave in the spin up state
\begin{equation}
\left\langle {{\mathbf{r}}}
 \mathrel{\left | {\vphantom {{\mathbf{r}} {{\psi _{{\text{ini}}}}}}}
 \right. \kern-\nulldelimiterspace}
 {{{\psi _{{\text{ini,a}}}}}} \right\rangle  = \exp \left( {i{\mathbf{p}} \cdot {\mathbf{r}}/\hbar} \right)\left( \begin{gathered}
  1 \hfill \\
  0 \hfill \\
\end{gathered}  \right),
\end{equation}
which is the eigenstate of $\mathcal{H}\left( {\kappa = 0,\gamma = 0 } \right)$ in the lower branch (we always consider positive $\alpha$). The target state for the ramping stage (a) is the corresponding eigenstate of $\mathcal{H}\left( {\kappa = 1,\gamma = 0 } \right)$:
\begin{equation}
\left\langle {{\mathbf{r}}}
 \mathrel{\left | {\vphantom {{\mathbf{r}} {{\psi _{{\text{tar}}}}}}}
 \right. \kern-\nulldelimiterspace}
 {{{\psi _{{\text{tar,a}}}}}} \right\rangle  = \exp \left( {i{\mathbf{p}} \cdot {\mathbf{r}}/\hbar} \right)\left( \begin{gathered}
  \alpha \hbar {\omega _0} + \sqrt {\frac{{9\hbar {\omega _{{\text{so}}}}{{\mathbf{p}}^2}}}{{2m}} + {{\left( {\alpha \hbar {\omega _0}} \right)}^2}}  \hfill \\
  3{\omega _{{\text{so}}}}({p_x} + i{p_y})/{k_{{\text{so}}}} \hfill \\
\end{gathered}  \right).
\end{equation}
Here the spin state is not normalized. We denote the wavefunction during stage (a) under the pulse sequence as $\left| {{\phi _{\mathbf{p}}}\left( t \right)} \right\rangle $. The fidelity between this state and the target state is then given by
\begin{equation}
  {F_{\mathbf{p}}}\left( t \right) = {\left| {\left\langle {{{\psi _{{\text{tar,a}}}}}}
 \mathrel{\left | {\vphantom {{{\psi _{{\text{tar, a}}}}} {{\phi _{\mathbf{p}}}\left( t \right)}}}
 \right. \kern-\nulldelimiterspace}
 {{{\phi _{\mathbf{p}}}\left( t \right)}} \right\rangle } \right|^2}.
\end{equation}
We notice that, the initial fidelity for stage (a) is
\begin{equation}
{F_{\mathbf{p}}}\left( {t = 0} \right) = {\left| {\left\langle {{{\psi _{{\text{tar,a}}}}}}
 \mathrel{\left | {\vphantom {{{\psi _{{\text{tar,a}}}}} {{\psi _{{\text{ini,a}}}}}}}
 \right. \kern-\nulldelimiterspace}
 {{{\psi _{{\text{ini,a}}}}}} \right\rangle } \right|^2} = \frac{{{{\left( {\alpha \hbar {\omega _0} + \sqrt {9\hbar {\omega _{{\text{so}}}}{{\mathbf{p}}^2}/2m + {{\left( {\alpha \hbar {\omega _0}} \right)}^2}} } \right)}^2}}}{{{{\left( {\alpha \hbar {\omega _0} + \sqrt {9{\hbar\omega _{{\text{so}}}}{{\mathbf{p}}^2}/2m + {{\left( {\alpha \hbar {\omega _0}} \right)}^2}} } \right)}^2} + 9{\hbar\omega _{{\text{so}}}}{{\mathbf{p}}^2}/2m}},
\end{equation}
which is close to unity for small momentums and large $\alpha{\omega _0}/\omega_{\text{so}}$. For instance, for the momentum point ${\mathbf{p}} = \left( {1,1} \right){\hbar k_{{\text{so}}}}$ in the lower energy branch, with $\omega_0 = 32.3\omega_{\rm so}=(2\pi)\times9.3\,\text{kHz}$ and $\alpha = 1$, we get ${F_{\mathbf{p}}}\left( {t = 0} \right) = 99.57\%$. After linearly ramping up $\kappa$ from zero to one in $N_a = 40$ cycles with period $T = 10\,\mu\text{s}$, the final fidelity is numerically found to be ${F_{\mathbf{p}}}\left( {t = {N_a}T} \right) = 1 - 4.3 \times {10^{ - 5}}$. This shows that the low energy eigenstates of $\mathcal{H}\left( {\kappa = 1,\gamma = 0 } \right)$ can be reached reliably.

\begin{figure}[hbtp]
  \includegraphics[width=0.45\columnwidth]{{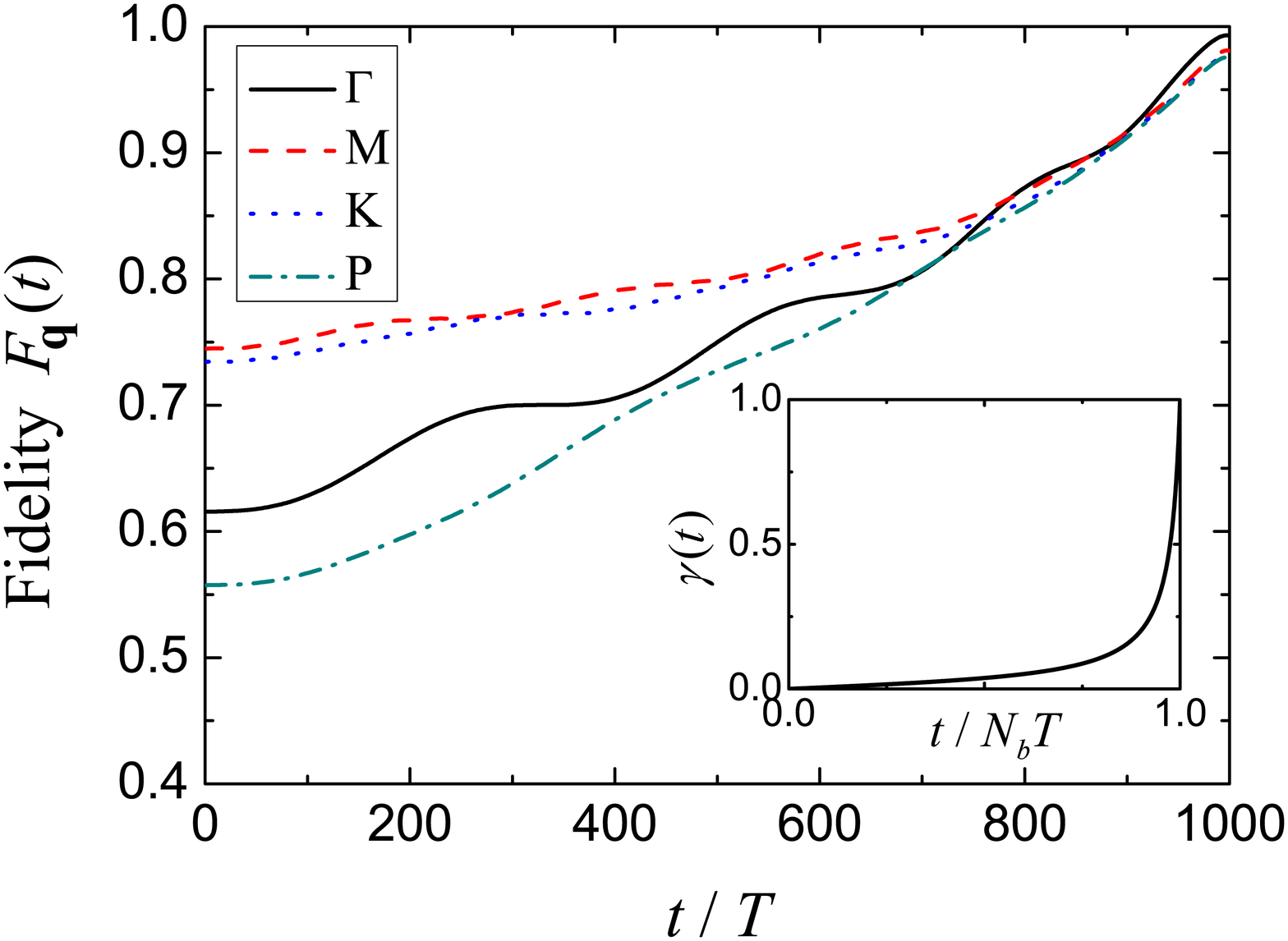}}\\
  \caption{The time-dependent fidelities for selected quasimomentum points in the first Brillouin zone. Here in the legend, P stands for the result for the quasimomentum point $\mathbf{q} = (0.3, -0.2){{k_{\text{so}}}}$. The strength of the magnetic lattice is ramped up as a tangent function (the inset) in $N_b = 1000$ cycles.  The parameters for the system are the following: $\omega_0 = 32.3\omega_{\rm so}=(2\pi)\times9.3\,\text{kHz}$, $\alpha = 1$ and $T = 10\,\mu\text{s}$.
  }\label{Fig_S4}
 \end{figure}

We then consider the ramping process for stage (b). We choose the initial state for this stage as the lowest eigenstates of $\mathcal{H}(\kappa=1, \gamma=10^{-6})$. The vanishingly small lattice term $10^{-6}\times\frac{1 }{8}\hbar {\omega _0}{\mathbf{M}} \cdot {\bm{\sigma }}$ introduced in our numerical simulation accounts for the effect of the Bragg reflections at the edges of the first Brillouin zone, which turns the good quantum number from momentum (denoted as $\mathbf{p}$) to quasimomentum (denoted as $\mathbf{q}$) restricted in the first Brillouin zone~\cite{Ashcroft1976}. The fidelity between the target state $\left| {{\psi _{1{\mathbf{q}}}}} \right\rangle $ [i.e., eigenstate of Eq.~(\ref{H_half_eff}) with the form of Eq.~(\ref{plane_wave_expansion_psi}) with band index $n=1$] and the time-dependent state $\left| {{\phi _{\mathbf{q}}}\left( t \right)} \right\rangle $ evolved by pulse sequence with increased magnetic lattice strength (which is proportional to $\gamma$) is defined as:
\begin{equation}
  {F_{\mathbf{q}}}\left( t \right) = {\left| {\left\langle {{{\psi _{ 1{\mathbf{q}}}}}}
 \mathrel{\left | {\vphantom {{{\psi _{n = 1,{\mathbf{q}}}}} {{\phi _{\mathbf{q}}}\left( t \right)}}}
 \right. \kern-\nulldelimiterspace}
 {{{\phi _{\mathbf{q}}}\left( t \right)}} \right\rangle } \right|^2}.
\end{equation}
We suggest ramping up the strength of the magnetic lattice as a tangent function (rather than a linear one), which corresponds to a time-dependent $\gamma$ function with the form
\begin{equation}\label{tan_ramp}
  \gamma \left( t \right) = \frac{{\tan \left[ {\eta \pi t/\left( {2{N_b}T} \right)} \right]}}{{\tan \left( {\eta \pi /2} \right)}},
\end{equation}
where the parameter $\eta $ is taken as $0.975$. The typical results for the time-dependent fidelity functions for different quasimomentums, together with the $\gamma \left( t \right)$ function, are shown in Fig.~\ref{Fig_S4}. As can be seen in this figure, the fidelities after ramping up the magnetic lattice approach unity for all tested quasimomentum points. We thus conclude that the eigenstates in the lowest energy band of the effective Hamiltonian ${H_{{\rm{eff}}}}$ in Eq.~(\ref{H_half_eff}) can be prepared appropriately through the adiabatic approach we propose.

\begin{thebibliography}{59}
\bibitem [{Vec()}]{VectorPotential2015}%
  \BibitemOpen
  \href@noop {} {}\bibinfo {note} {The vector potential $\bf A$ contains a
  gauge-dependent singularity, which is chosen to be at the center of the unit
  cell for the gauge we use.}\BibitemShut {Stop}%
\end{thebibliography}
\end{document}